# Designing Knowledge Tools: How Students Transition from Using to Creating Generative AI in STEAM classroom

Qian Huang
*Lee Kuan Yew Center for Innovative Cities*
*Singapore University of Technology and Design*
Singapore, Singapore
qian_huang@sutd.edu.sg

Nachamma Sockalingam
*Office of Strategic Planning*
*Singapore University of Technology and Design*
Singapore, Singapore
nachamma@sutd.edu.sg

Thijs Willems
*Lee Kuan Yew Center for Innovative Cities*
*Singapore University of Technology and Design*
Singapore, Singapore
thijs_willems@sutd.edu.sg

King Wang Poon
*Lee Kuan Yew Center for Innovative Cities*
*Singapore University of Technology and Design*
Singapore, Singapore
poonkingwang@sutd.edu.sg

*Abstract*—This study explores how graduate students in an urban planning program transitioned from passive users of generative AI to active creators of custom GPT-based knowledge tools. Drawing on Self-Determination Theory (SDT), which emphasizes the psychological needs of autonomy, competence, and relatedness as foundations for intrinsic motivation, the research investigates how the act of designing AI tools influences students' learning experiences, identity formation, and engagement with knowledge. The study is situated within a two-term curriculum, where students first used instructor-created GPTs to support qualitative research tasks and later redesigned these tools to create their own custom applications, including the Interview Companion GPT. Using qualitative thematic analysis of student slide presentations and focus group interviews, the findings highlight a marked transformation in students' roles and mindsets. Students reported feeling more autonomous as they chose the functionality, design, and purpose of their tools, more competent through the acquisition of AI-related skills such as prompt engineering and iterative testing, and more connected to peers through team collaboration and a shared sense of purpose. The study contributes to a growing body of evidence that student agency can be powerfully activated when learners are invited to co-design the very technologies they use. The shift from AI tool users to AI tool designers reconfigures students' relationships with technology and knowledge, transforming them from consumers into co-creators in an evolving educational landscape.

## I. Introduction

Generative Artificial Intelligence (AI) has rapidly emerged as a powerful new medium for knowledge work in education. Tools like ChatGPT can serve as knowledge tools for instructors and students, providing on-demand information and even creative collaboration [1]- [4]. In higher education, students have begun to adopt generative AI for tasks ranging from research assistance to writing support [5]. However, the typical use case positions students as end-users consuming AI-generated knowledge rather than as creators of AI tools themselves. This paper explores a novel educational approach in an urban planning course where students transitioned from simply using generative AI to designing their own AI-driven knowledge tools. By building custom GPT-based applications tailored to planning tasks, students took on the role of co-creators of knowledge technology. We investigate how this hands-on creation process influenced their learning experience, particularly in terms of student agency and ownership of knowledge.

Custom GPTs are specialized instances of large language models (LLMs), such as OpenAI's GPT-4, that have been tailored to specific tasks, domains, or user needs. Unlike the general GPT, which is trained on a broad corpus of data, a Custom GPT is configured through additional instructions, curated datasets, or integrations with external tools to refine its behavior, tone, and knowledge boundaries. By creating a Custom GPT, users can build AI assistants that perform domain-specific reasoning (e.g., summarizing legal texts, simulating interviews, or assisting in qualitative coding), while retaining the general language understanding and generation abilities of the base GPT model. We investigate how this hands-on creation process influenced their learning experience, particularly in terms of student agency and ownership of knowledge.

There is a growing recognition that actively engaging with generative AI can deepen learning. Recent studies suggest that higher-order learning outcomes occur when students use generative AI to construct and augment knowledge, as opposed to using it passively or procedurally [6]- [7]. Building on this insight, our study positions students as developers of AI tools: each student team in a graduate urban planning course was tasked with designing a custom GPT-based application to address a real-world planning or learning challenge. Through this process, students were not only learning about AI but also embedding their own domain knowledge into AI, effectively taking ownership of the knowledge the AI would dispense. This approach aligns with student-centered and constructionist learning theories [8]- [9] that emphasize the value of learners creating tangible artifacts as part of knowledge construction.

Our investigation focuses on how this creator experience affected students' sense of autonomy, competence, and relatedness in learning. We adopt Self-Determination Theory (SDT) as a lens to analyze student experiences, as it centers on these three psychological needs as fundamental to intrinsic motivation [10]. The context for the study is an urban planning education setting, where technological innovation (like AI) is increasingly relevant. By examining student reflections and interviews after completing the course, we shed light on the transition from user to creator of AI and its pedagogical implications. The findings highlight that when students became AI tool designers, they developed a stronger sense of agency (making choices about what their AI would do), gained confidence and skills in using AI critically, and collaborated meaningfully with peers, all of which point to a more empowered learning experience.

## II. Literature

The integration of generative AI in various education has been met with both excitement and concern [5],[11]. On one hand, AI tools promise to enhance learning by providing personalized assistance, rapid information retrieval, and new forms of creative engagement [12]- [23]. In fields like urban planning, scholars have begun to view AI as a potential "creative partner" in the design process, suggesting that techniques such as prompt engineering could become analogous to design thinking in the planner's toolkit [14]. Early classroom applications of ChatGPT and similar models have largely treated the technology as a tool for students to use, for example, to generate ideas, summarize readings, or draft essays [15]. These uses, while convenient, risk positioning students as passive recipients of AI outputs. Educators have raised questions about how such use might affect critical thinking, originality, and the student's sense of authorship over their work [4], [16].

In response, some researchers argue that generative AI should be harnessed in ways that promote active learning and student agency, rather than replacing student effort [17]. For instance, a recent study found that when students engaged with GenAI in a constructive manner (using it to build on their knowledge), they achieved deeper learning than those who used it in a surface-level, copy-paste fashion [6]. This suggests that pedagogical designs should move students from consumers of AI to collaborators or even creators, thereby encouraging them to critically engage with AI outputs and underlying knowledge.

Student agency in learning technology integration is a key theme in the literature. Agency refers to students' capacity to make meaningful choices in their learning process and to contribute to knowledge creation [18]. Prior work on knowledge-building pedagogies has emphasized the importance of treating students as contributors to a collective knowledge base rather than just learners of established facts. In knowledge-building classrooms [9], students create and improve ideas collaboratively, which inherently gives them ownership of knowledge. Generative AI, if used naively, could undermine this by providing ready answers and obscuring sources, thus "reducing many voices to one" authoritative voice of the AI [19]. However, when students are tasked with creating their own AI-driven knowledge tools, the dynamic shifts—they must decide what content to include, how the AI should function, and for what purpose. This can enhance their sense of responsibility for the knowledge the AI provides. It also aligns with constructionist approaches [6] where learners gain understanding through building artifacts that have meaning in the real world.

Another relevant strand of literature concerns AI literacy and empowerment. As AI becomes ubiquitous, scholars argue students should not only learn how to use AI tools but also understand how they work and how to create or customize them [20]. Empowering students to create AI applications can demystify the "black box" of AI and develop critical digital skills. Early efforts in K–12 have shown that when students are guided to leverage AI for their own projects, they develop a nuanced understanding of AI's capabilities and limitations and become more critical and mindful users of these technologies [21]. This is especially pertinent in higher education and professional fields like urban planning, where tomorrow's practitioners will likely collaborate with AI in their work. By designing AI tools, students in our study directly engaged with questions of data, model output quality, and user interaction, thereby gaining AI literacy in addition to domain knowledge.

Urban planning education is increasingly exploring technology integration, with GIS, data analytics, and now AI becoming part of the curriculum. Yet, little documented research exists on students in planning (or similar design fields) actively creating AI-based tools. Our study contributes to filling this gap by providing an empirical account of graduate students designing generative AI applications in an urban planning context. This unique setting combines technical learning with a strong emphasis on human-centric design, as planning students had to ensure their AI tools would be useful for people (e.g., fellow students or practitioners). The literature on human-computer interaction and design thinking in education suggests that having students think about end-users can deepen their empathy and sense of purpose, which relates to social aspects of learning [22]-[23]. We anticipated similar outcomes here: that in building tools intended for real users (even if hypothetical), students would feel a greater connection to a community of users and thus find the project more meaningful.

In summary, existing research underscores the importance of engaging students as active participants in the age of generative AI. Using AI to support knowledge construction and creativity tends to yield better learning outcomes than using it for passive consumption. Additionally, granting students ownership in the creation of technological tools is aligned with fostering agency, critical thinking, and deeper learning. Our work extends these ideas by examining a concrete instance where students moved from using an AI tool to building one, and by analyzing the experience through the theoretical lens of Self-Determination Theory to understand its impact on their motivation and learning mindset.

### III. THEORETICAL FRAMEWORK

To analyze the students' transition from AI users to AI creators, we employ Self-Determination Theory (SDT) as our theoretical framework. SDT, formulated by Deci and Ryan [24]-[25], is a theory of human motivation that identifies three innate psychological needs autonomy, competence, and relatedness as essential for fostering intrinsic motivation and engagement in any activity. When these needs are satisfied, individuals are more likely to feel self-motivated, take initiative, and persist in their efforts; if the needs are thwarted, motivation and well-being can suffer [25]. In the context of learning, SDT provides a useful lens to examine how certain educational experiences either support or hinder students' psychological needs and thereby affect their motivation and growth [26].

Autonomy in SDT refers to feeling that one's actions are self-chosen and aligned with one's own interests and values. It is the sense of volition and ownership over one's behavior [25] In a learning environment, students experience autonomy when they have meaningful choices, the freedom to explore their own ideas, and the ability to shape their learning process. Competence denotes the feeling of effectiveness and mastery in one's activities. A student who feels competent believes they can meet the challenges given to them, understands the material, and sees progress in their skills [25]. Relatedness is the need to feel connected to others, to have a sense of belonging, mutual respect, and that one's contributions matter to a community. In an academic setting, relatedness can come from collaboration with peers, positive interactions with instructors, or knowing that one's work benefits others. SDT posits that a well-designed learning experience should support all three needs: give students some autonomy, optimally challenge them to build competence, and provide social support and relevance to satisfy relatedness [25].

We chose SDT to frame our analysis because the act of creating a generative AI tool in a course is expected to touch on all three of these dimensions of motivation. First, autonomy: Unlike typical assignments with narrowly defined instructions, designing an AI tool gave students a high degree of choice, they decided on the tool's purpose, content, and functionality. This creative freedom and the shift of role (from following a tool's outputs to determining what the tool should do) likely enhanced their sense of control over the learning activity. We anticipated that students would feel a greater ownership of their learning outcomes, as they were literally building the instrument of their learning. Second, competence: Building a functional AI-based application is a challenging task, especially for students with no prior programming experience (as was the case for many in this class). Through this project, students had to acquire new skills such as prompt engineering, curation of a knowledge base for the AI, and iterative testing and refining of the AI's performance. By overcoming these challenges and seeing a working prototype of their custom GPT, students could gain a strong sense of efficacy and mastery. We expected statements in their reflections indicating increased confidence in using AI or pride in the technical skills gained. Additionally, by better understanding how the AI works "behind the scenes," students would feel more competent in using such tools critically in the future. Third, relatedness: The project was inherently collaborative. Students worked in teams to build their tools, and the class as a whole formed a community exploring generative AI. We anticipated that this prosocial aspect and teamwork would fulfill students' need for relatedness. They might feel more connected to classmates through shared problem-solving, and find meaning in creating something that could benefit their fellow students or the broader community. SDT highlights that supporting relatedness can involve showing the relevance of an activity to one's social world and providing a sense of contributing to something larger than oneself, our project's design encouraged such connections.

In summary, SDT provides a structured way to interpret the students' experiences by examining evidence of autonomy, competence, and relatedness need satisfaction. When analyzing the qualitative data from student interviews and reflections, we specifically looked for expressions that indicated how the project influenced their sense of freedom and agency in learning (autonomy), their skills and understanding (competence), and their interpersonal or community connections (relatedness). The following sections (Findings) are organized around these three needs, illustrating how the transition from user to creator of AI tools impacted each aspect of the students' motivational and learning experience.

### IV. METHODOLOGY

This study was conducted in a Master's degree program in Urban Science, Policy and Planning (MUSPP) involving 17 students. The course structure was organized across two terms. In Term 1, two instructors introduced the students to qualitative research methods and built four custom GPTs to enhance student learning, including custom GPTs for training qualitative research knowledge, observation skills, interview skills and design thinking skills. These GPTs were designed to train students' skills in qualitative observation, interview techniques, field note-taking, and writing. Integrating AI-based tools into classroom settings is increasingly recognized as a method for promoting active learning and enhancing applied research skills.

In Term 2, the same student cohort continued under two different instructors in a *Creating the Frontiers of the No-code Smart City* Class. Here, students were challenged not just to use but to create their own custom GPTs by improving two of the original GPTs from Term 1: the Observation Station GPT and the Interview Simulation GPT. This approach reflects broader educational trends toward involving students as co-designers in technology-enhanced learning environments.

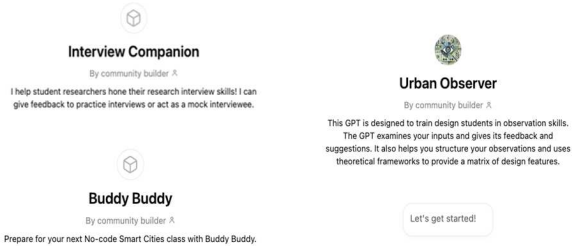

Fig. 1. Custom GPTs created by students teams in Term 2.

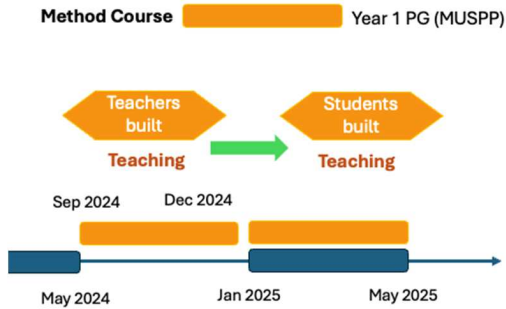

Fig. 2. Transtion from using Gen-AI tools to creating Gen-AI tools by students in one academic year in MUSPP program.

The focal point for this paper is on the Interview Simulation GPT, which students developed into what they named the Interview Companion GPT. Students in Term 2 reflected on the pros and cons of the Interview Simulation GPT, identified areas for improvement, and designed enhanced features based on their learning needs.

*A. Data Collection*

The research team gathered qualitative data from several sources. First, students created slide presentations documenting their process of designing custom GPTs. These slides included their reflections on the strengths and weaknesses of both the Term 1 GPT and their improved version, offering insight into their experiences, perceptions, and learning outcomes. Collecting self-reflective artifacts such as presentation slides is a well-established method in educational technology research, supporting deeper understanding of student agency and design thinking processes.

Second, semi-structured focus group interviews were conducted with students. Four transcripts were collected and analyzed. These interviews explored students' experiences using the GPTs in Term 1 and creating GPTs in Term 2, focusing on their sense of learning agency, challenges encountered, collaborative experiences, and the knowledge ownership process. Semi-structured interviews are widely used in qualitative educational research to capture complex perceptions and motivations, particularly when exploring technology adoption and learning experiences [29]-[30]. The study was reviewed and approved by the university's Institutional Review Board (IRB), and the students signed a consent form. In this article, pseudonyms were used when quoting students directly, except in Figure 3, where the student's real name was used to give credit.

*B. Data Analysis*

A qualitative thematic analysis was employed, grounded in Self-Determination Theory (SDT). The research team coded interview transcripts and slide content to identify themes corresponding to the three core psychological needs outlined by SDT: autonomy, competence, and relatedness. Thematic analysis provides a structured yet flexible approach for analyzing patterns in qualitative data while maintaining fidelity to theoretical frameworks such as SDT.

V. FINDING

The findings from this study are organized according to the three core psychological needs outlined by Self-Determination Theory: autonomy, competence, and relatedness. Each subsection integrates student reflections, focus group interviews, and the case study of the Interview Companion GPT.

*A. Autonomy: From Users to Creators*

Students consistently described a significant shift in their learning experience when transitioning from users to creators of generative AI tools. In Term 1, students used the custom GPTs instructors built. In Term 2, by improving and redesigning these GPTs, students experienced a strong sense of autonomy. Students chose what functionalities to add based on their needs and reflections.

> *Our GPT provides live feedback plus suggestions based on text, audio, Term 1 GPT only text interviews. (Lucy)*

The learning environment promoted autonomy not only through tool customization but also through open-ended problem-solving. Students expressed feeling empowered and critical to decide the purpose, features, and content of their GPTs.

> *Learning how to create it makes me more conscious about how to be a good user of it... knowing the back end of it makes me more critical about what I see as a front-end user. (Steve)*

In the process of creating their own custom GPTs, students have the autonomy to take ownership of and apply their knowledge.

> *Our idea was also coming from our past experienes…it was not some case study or some examples that we are putting in, but what human experience that we have to put in…That's a part of process. (Alex)*

> *I think that the experience of building a custom one meant that you had to apply a lot of intentionality to its use, because you need to ask yourself questions about translating some kind of use, some kind of design. (May)*

B. *Competence: Gaining Skills and Confidence*

The experience of creating and designing GPTs enhanced students' competence in several areas. Technically, students learned prompt engineering, prompt refining, and the configuration of AI tools. For instance, they developed multiple feedback modes in their own custom GPTs.

Students reported gaining not only technical skills but also confidence in using AI tools critically. Students shifted from passive acceptance of AI outputs to critical engagement and refinement.

> *When I was part of the creation process, it helped me understand what it can do and what the limitations are, and that awareness made me more open to using it. (Lily)*

During the creation process, students tested multilingual capabilities and feedback functions. They observed strengths, such as accurate language switching, and weaknesses, such as unnatural accents and overly polite feedback. These reflections showed a deep understanding of the AI tool's capabilities and limitations.

> *At the end of it, I think I really gained a lot of knowledge and ksills through the hands-on activities. (Max)*

C. *Relatedness: Collaborative Creation and Community*

The process of creating custom GPTs fostered a strong sense of relatedness among students. Collaboration was essential: students worked in teams, shared feedback, and tested each other's GPTs.

> *There were certain refinements, certain testing that we do, even among classmates, in order to further refine and make the custom GPT sharper and fit for purpose. (Mike)*

> *I think teamwork is about adding on my ideas, questioning my ideas (during the process of creating GPT). (Linda)*

Beyond teamwork, students felt connected to the broader learning community. Many expressed a sense of contributing to future cohorts, as their GPTs could benefit others.

> *Maybe our bot can tutor junior students next term – that'd be so cool, like leaving behind a legacy! (Steve)*

This reflects a meaningful connection between students' work and the wider academic environment.

D. *Case Study: Interview Companion GPT*

The Interview Companion GPT project exemplifies how student-created generative AI tools embody the principles of autonomy, competence, and relatedness outlined in our broader findings. Developed by a team of six students in Term 2, this custom GPT was an iterative enhancement of the Interview Simulation GPT originally introduced by instructors in Term 1. The student team began by critically analyzing the limitations of the initial tool. While the original GPT could simulate text-based interview responses, students identified several key gaps: it lacked live feedback functionality, could not handle multilingual interactions effectively, and provided only generic responses that did not reflect real-world interview complexity. Recognizing these constraints, students collaboratively designed and implemented new features aimed at improving both the learning experience and the tool's practical utility.

The Interview Companion GPT incorporated four main functions:

- Feedback on Text Transcripts: Students enabled the GPT to provide structured feedback based on uploaded text transcripts of interviews, allowing users to review and refine their performance post-interview.
- Live Audio Feedback During Mock Interviews: Using laptops or phones, the GPT could now listen to live interview sessions and deliver real-time feedback, simulating a more authentic interview environment. This represented a significant technical leap from static text responses.
- Mock Interviewee Role: The GPT could act as an interviewee, responding in real-time to user questions while simultaneously evaluating user performance based on conversational tone, pacing, and content.
- Video Feature for Live Vision Analysis: Though conceptualized, this function, intended to analyze interviewees' body language and facial expressions, was not implemented due to current technical limitations on the platform.

Throughout the project, students documented both their design rationale and challenges encountered. It shows that while the tool was not perfect, it provided valuable opportunities for self-assessment and iterative learning.

> *Comparison of mock vs. real interviews… generated comparison table on tone, pacing & answers. Not accurate but allowed performance reflection.(Lucy)*

Another student expressed a forward-looking and high self-efficacy attitude: "*Hopefully, this can become Research Interview Simulator 2.0!!*" suggesting an awareness of continuous improvement as an integral part of AI tool development.

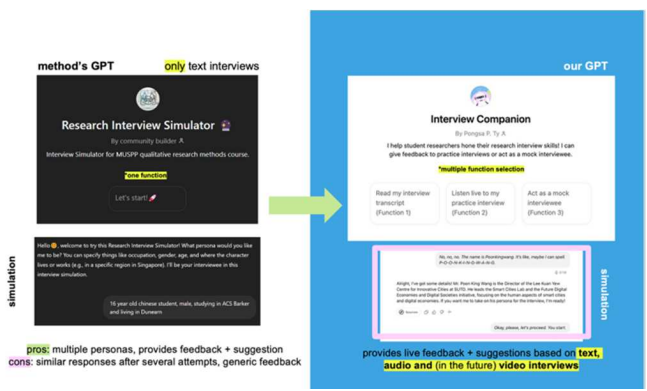

Fig. 3. Students description of their transition (The Interview Custom GPT Team with real names: Baorong Li, Calvin Tan Wei Ru, Cherie Lynne Gomintong, Jose Paolo Tambagan Boquiren, Nur Dini Binti Mahmud, Ty Puthipongsa)

At the meantime, critical reflections also emerged regarding several shortcomings:

- Accent and Cultural Nuance: *"AI's accent sounded unnatural, lacks cultural nuance."* It shows reflects sensitivity to the limitations of AI-generated speech and the importance of human-like interaction in interview simulations.
- Feedback Quality: "*Feedback lacks criticality… in real interviews, interviewers give tough/ambiguous questions.*" Students recognized that while the GPT could provide surface-level feedback, it fell short in replicating the nuanced and challenging nature of real-world interviews.
- Technical Reliability: "*Crashes and pauses due to Wi-Fi connection issues.*" It highlighted the dependency of cloud-based AI tools on stable internet connectivity, affecting usability in real-time scenarios.

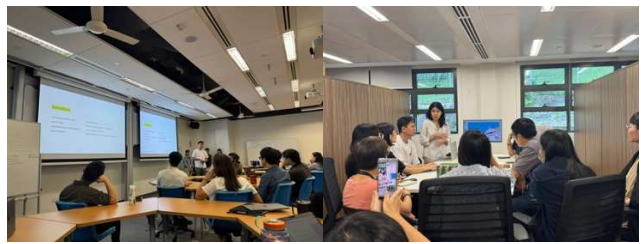

Fig. 4. The Interview Custom GPT Team presented in *Creating the Frontiers of the No-code Smart City* Class (left) and at the Oral History Center National Archieves in Singapore (right). (https://urbanscience.sutd.edu.sg/news-and-events/muspp-24-students-shine-at-the-oral-history-centre-national-archives-with-customgpt-interview-companion/)

These insights were captured through both slide presentations and focus group interviews, offering rich qualitative evidence of students' critical engagement with their project. The process of evaluating their own work through reflective practice not only strengthened their technical skills but also cultivated a mindset of continuous learning and tool refinement. Ultimately, the Interview Companion GPT case study illustrates how giving students ownership over AI tool development fosters deeper learning, practical problem-solving abilities, and an authentic sense of agency in navigating emerging technologies.

## VI. DISCUSSION

The findings of this study highlight several important considerations for educators and researchers interested in integrating generative AI creation into higher education. While the structure and results align closely with Self-Determination Theory (SDT), the students' experiences reveal additional layers of insight into how ownership of AI tool development influences learning, motivation, and educational identity [25].

*A. Transitioning Roles: From Passive Users to Knowledge Designers*

One of the clearest patterns emerging from our analysis is the transformational effect of shifting students' roles from users to creators. As reflected in both student quotes and the Interview Companion GPT case study, creating a custom GPT catalyzed a mindset shift. Students described becoming more intentional, critical, and self-directed. This process aligns with constructionist learning theories, which emphasize learning through building tangible artifacts [8]. It also reinforces SDT's concept of autonomy: students were not merely completing a predefined task but determining the direction, content, and impact of their own tools. This level of control distinguishes generative AI creation from other forms of technology-enhanced learning, where the tools themselves remain fixed and outside the learner's influence [9].

*B. Learning Through Reflection and Iteration*

A notable outcome is the depth of learning that occurred through iterative design and reflection. Students reported that building a GPT required them to articulate exactly what knowledge was valuable, how it should be structured, and how users might interact with it. This reflective process mirrors what the terms "the reflective practitioner" model, wherein professional learning occurs through cycles of action and reflection [28]. In our context, students not only learned about AI but also deepened their understanding of urban planning concepts by encoding them into the GPT. The act of "teaching the AI" became a form of learning-by-teaching, enhancing both technical competence and domain knowledge [29].

*C. Balancing Autonomy with Structure: The Role of Instructional Scaffolding*

While students valued their autonomy, many also noted the importance of instructor guidance. Several participants explicitly mentioned how frameworks provided by instructors helped them navigate initial challenges, especially in understanding AI's affordances and limitations. This suggests that while autonomy is crucial, it must be balanced with well-timed instructional support. Without such scaffolding, students might experience frustration or disengagement. This balance aligns with Vygotsky's concept of the "zone of proximal development," where learners perform best when supported just beyond their independent capability level [30].

*D. Research Limitations and Future Research*

This study is limited by its focus on a single cohort within one Master's program, which may constrain the generalizability of findings to other disciplines or institutional contexts. Additionally, the evaluation relied primarily on self-reported reflections and interviews; objective performance measures were not collected. Future research will extend this work into Term 3, where students undertake their Master's Research Project (MRP). Researchers will investigate whether and how students apply the knowledge and skills gained from using and creating custom GPTs to real-world urban planning challenges in their MRPs. This next phase will provide longitudinal insights into the transferability and sustained impact of AI literacy and tool creation experiences on professional-level projects.

VII. CONCLUSION

This study explored how graduate students in an urban planning program transitioned from using generative AI tools to creating their own, focusing on the development of custom GPTs in a classroom setting. Through the lens of Self-Determination Theory (SDT), we found that students' psychological needs for autonomy, competence, and relatedness were strongly supported throughout this process. By moving from users to creators of generative AI, students experienced greater ownership of knowledge and enhanced learning agency.

The shift to creating GPTs allowed students to make meaningful choices about tool functionality and design, fostering autonomy. It also challenged them to acquire new skills, such as prompt engineering and iterative design, thereby increasing their competence. Finally, the collaborative nature of the projects and the intention to create tools that benefit others cultivated a sense of relatedness among students and with the broader academic community.

The case study of the Interview Companion GPT exemplifies these dynamics in action. Students not only identified and addressed limitations in an instructor-built GPT but also envisioned further improvements, such as multilingual capabilities and live feedback functions. Their reflections demonstrated both critical awareness and a proactive attitude toward continuous development.

While this study focused on a specific educational context, its findings have broader implications for curriculum design in higher education. Encouraging students to create their own AI tools can demystify complex technologies, foster critical digital literacy, and enhance engagement through increased agency. Educators might consider integrating similar projects across disciplines, using accessible no-code platforms to lower technical barriers.

In sum, designing generative AI tools in educational settings does more than teach students about AI. It reshapes their relationship to knowledge itself. By giving students the tools and the mandate to create, educators can cultivate not only more skilled professionals but also more autonomous, competent, and socially connected learners.

## ACKNOWLEDGMENT

This research is supported by the Ministry of Education, Singapore, under its Tertiary Education Research Fund (MOE2024-TRF-30). Any opinions, findings and conclusions or recommendations expressed in this material are those of the author(s) and do not reflect the views of the Ministry of Education, Singapore.